\PassOptionsToPackage{unicode}{hyperref}
\PassOptionsToPackage{hyphens}{url}
\PassOptionsToPackage{dvipsnames,svgnames,x11names}{xcolor}
\documentclass[
  11pt,
  english,
  letterpaper,
]{article}
\usepackage{amsmath,amssymb}
\usepackage{lmodern}
\usepackage{iftex}
\ifPDFTeX
  \usepackage[T1]{fontenc}
  \usepackage[utf8]{inputenc}
  \usepackage{textcomp} 
\else 
  \usepackage{unicode-math}
  \defaultfontfeatures{Scale=MatchLowercase}
  \defaultfontfeatures[\rmfamily]{Ligatures=TeX,Scale=1}
\fi
\IfFileExists{upquote.sty}{\usepackage{upquote}}{}
\IfFileExists{microtype.sty}{
  \usepackage[]{microtype}
  \UseMicrotypeSet[protrusion]{basicmath} 
}{}
\makeatletter
\@ifundefined{KOMAClassName}{
  \IfFileExists{parskip.sty}{%
    \usepackage{parskip}
  }{
    \setlength{\parindent}{0pt}
    \setlength{\parskip}{6pt plus 2pt minus 1pt}}
}{
  \KOMAoptions{parskip=half}}
\makeatother
\usepackage{xcolor}
\IfFileExists{xurl.sty}{\usepackage{xurl}}{} 
\IfFileExists{bookmark.sty}{\usepackage{bookmark}}{\usepackage{hyperref}}
\hypersetup{
  pdftitle={Estimating the timing of stillbirths in countries worldwide using a Bayesian hierarchical penalized splines regression model},
  pdflang={en},
  colorlinks=true,
  linkcolor={blue},
  filecolor={Maroon},
  citecolor={blue},
  urlcolor={blue},
  pdfcreator={LaTeX via pandoc}}
\usepackage[margin=25mm]{geometry}
\usepackage{longtable,booktabs,array}
\usepackage{calc} 
\usepackage{etoolbox}
\makeatletter
\patchcmd\longtable{\par}{\if@noskipsec\mbox{}\fi\par}{}{}
\makeatother
\usepackage{footnote} 
\makesavenoteenv{longtable}
\usepackage{graphicx}
\makeatletter
\def\maxwidth{\ifdim\Gin@nat@width>\linewidth\linewidth\else\Gin@nat@width\fi}
\def\maxheight{\ifdim\Gin@nat@height>\textheight\textheight\else\Gin@nat@height\fi}
\makeatother
\setkeys{Gin}{width=\maxwidth,height=\maxheight,keepaspectratio}
\makeatletter
\def\fps@figure{htbp}
\makeatother
\newlength{\cslhangindent}
\newlength{\csllabelwidth}
\newlength{\cslentryspacingunit} 
\newenvironment{CSLReferences}[2] 
 {
  \setlength{\parindent}{0pt}
  \ifodd #1
  \let\oldpar\par
  \def\par{\hangindent=\cslhangindent\oldpar}
  \fi
  \setlength{\parskip}{#2\cslentryspacingunit}
 }%
 {}
\usepackage{calc}

\usepackage{float}
\usepackage{colortbl}
\usepackage{pdflscape}
\usepackage{tabu}
\usepackage{threeparttable}

\makeatletter
\newcommand\keywordsname{Keywords}

\makeatother

\usepackage{booktabs}
\usepackage{longtable}
\usepackage{array}
\usepackage{multirow}
\usepackage{wrapfig}
\usepackage{float}
\usepackage{colortbl}
\usepackage{pdflscape}
\usepackage{tabu}
\usepackage{threeparttable}
\usepackage{threeparttablex}
\usepackage[normalem]{ulem}
\usepackage{makecell}
\usepackage{xcolor}
\usepackage{subfig}
\usepackage{amsmath}
\usepackage{setspace}
\usepackage{booktabs}
\usepackage{longtable}
\usepackage{array}
\usepackage{multirow}
\usepackage{wrapfig}
\usepackage{float}
\usepackage{colortbl}
\usepackage{pdflscape}
\usepackage{tabu}
\usepackage{threeparttable}
\usepackage{threeparttablex}
\usepackage[normalem]{ulem}
\usepackage{makecell}
\usepackage{xcolor}
\ifXeTeX
  \usepackage{polyglossia}
  \setmainlanguage[]{english}
  \setotherlanguage[]{english}
\else
  \usepackage[english,main=english]{babel}

\def\languageshorthands#1{}
  
  \newenvironment{english}[2][]{\begin{otherlanguage}{english}}{\end{otherlanguage}}
\fi
\ifLuaTeX
  \usepackage{selnolig}  
\fi

\title{Estimating the timing of stillbirths in countries worldwide using a Bayesian hierarchical penalized splines regression model}

\makeatletter
\def\MyNewLabel#1#2#3{\expandafter\gdef\csname #1@#2\endcsname{#3}}

\def\MyRef#1#2{\@ifundefined{#1@#2}{???}{\csname #1@#2\endcsname}}

\newcommand*\ifcounter[1]{%
  \ifcsname c@#1\endcsname
    \expandafter\@firstoftwo
  \else
    \expandafter\@secondoftwo
  \fi
}
\makeatother

\MyNewLabel{ADDRTXT}{A}{Department of Statistical Sciences, University of Toronto}
\MyNewLabel{ADDRTXT}{B}{Departments of Statistical Sciences and Sociology, University of Toronto}

\MyNewLabel{ANOTETXT}{corresp}{\href{mailto:myc.chong@mail.utoronto.ca}{\nolinkurl{myc.chong@mail.utoronto.ca}}. The authors would like to acknowledge Lucia Hug, Danzhen You, David Sharrow, Lucy Smith, Hannah Blencowe, Leontine Alkema, Zhengfan Wang, and Jon Wakefield for their invaluble input and feedback.}

\usepackage{bigfoot}
\DeclareNewFootnote{Addr}[arabic] 
\DeclareNewFootnote{ANote}[fnsymbol]

\usepackage{authblk}

\newcounter{addrcnt}

\makeatletter
\newcommand*\createaddrlblbycode[1]{%
  \ifcounter{ADDRLBL@#1}
    {}
    {\refstepcounter{addrcnt}\newcounter{ADDRLBL@#1}\setcounter{ADDRLBL@#1}{\value{addrcnt}}}%
}

\newcommand*\addrlblbycode[1]{\arabic{ADDRLBL@#1}}

\newcommand*\addrbycode[1]{%
  \ifcounter{ADDR@#1}
    {}
    {\newcounter{ADDR@#1}%
     \affil[\addrlblbycode{#1}]{\MyRef{ADDRTXT}{#1}}}%
}

\newcommand*\createanotelblbycode[1]{%
  \ifcounter{ANOTELBL@#1}
    {}
    {\refstepcounter{footnoteANote}\newcounter{ANOTELBL@#1}\setcounter{ANOTELBL@#1}{\value{footnoteANote}}}%
}

\newcommand*\anotelblbycode[1]{\fnsymbol{ANOTELBL@#1}}

\newcommand*\anotebycode[1]{%
  \ifcounter{ANOTE@#1}
    {}
    {\newcounter{ANOTE@#1}%
     \footnotetextANote[\value{ANOTELBL@#1}]{\MyRef{ANOTETXT}{#1}}}%
}
\makeatother

\createaddrlblbycode{A}

\createanotelblbycode{corresp}

\author[%
\addrlblbycode{A}%
,%
$\anotelblbycode{corresp}$%
]{Michael Y.C. Chong}

\addrbycode{A}

\createaddrlblbycode{B}

\author[%
\addrlblbycode{B}%
]{Monica Alexander}

\addrbycode{B}

\date{}

\begin{document}
\maketitle

\anotebycode{corresp}


\begin{english}

\begin{abstract}
Reducing the global burden of stillbirths is important to improving child and maternal health. Of interest is understanding patterns in the timing of stillbirths --- that is, whether they occur in the intra- or antepartum period --- because stillbirths that occur intrapartum are largely preventable. However, data availability on the timing of stillbirths is highly variable across the world, with low- and middle-income countries generally having few reliable observations. In this paper we develop a Bayesian penalized splines regression framework to estimate the proportion of stillbirths that are intrapartum for all countries worldwide. The model accounts for known relationships with neonatal mortality, pools information across geographic regions, incorporates different errors based on data attributes, and allows for data-driven temporal trends. A weighting procedure is proposed to account for unrepresentative subnational data. Results suggest that the intrapartum proportion is generally decreasing over time, but progress is slower in some regions, particularly Sub-Saharan Africa.

\end{abstract}

\end{english}

\newpage

\hypertarget{introduction}{%
\section{Introduction}\label{introduction}}

Stillbirths represent a large share of the global burden on child and maternal health. Globally, in 2019 there were an estimated 2 million stillbirths (babies born with no sign of life at 28 weeks of pregnancy or later) (\protect\hyperlink{ref-hug2021global}{Hug et al. 2021}). This represents a rate of around 14 stillbirths per 1000 total births, which is of comparable magnitude to the global neonatal mortality rate of around 17 deaths per 1000 live births (\protect\hyperlink{ref-hug2019national}{Hug et al. 2019}). As such, there have been increased calls to monitor stillbirth outcomes in addition to infant and maternal deaths to fully understand the extent of the risks faced by different populations during pregnancy and childbirth (\protect\hyperlink{ref-lawn2016stillbirths}{Lawn et al. 2016}). The Every Newborn Action Plan (ENAP), which was
endorsed by 194 WHO Member States, calls for each
country to achieve a rate of 12 stillbirths or fewer per
1000 total births by 2030 (\protect\hyperlink{ref-world2014every}{World Health Organization 2014}). While progress has been made in reducing stillbirths, substantial inequalities across countries and regions still persist (\protect\hyperlink{ref-hug2020neglected}{Hug et al. 2020}).

An important part of working towards the goal of ending preventable stillbirths is having reliable estimates of when stillbirths occur, that is, whether a stillbirth occurs before or after the onset of labor (ante- versus intrapartum). Stillbirths that occur intrapartum are likely to be preventable in most situations given adequate access to healthcare (\protect\hyperlink{ref-de2016stillbirths}{De Bernis et al. 2016}), and thus are likely to be responsive to increased resources and health policy changes. The goal of this project is thus to estimate the proportion of stillbirths that are intrapartum for all countries and regions worldwide, over the period 2000-2021.

Challenges in estimation exist, particularly in low- and middle-income countries, due to data availability issues. Even when data do exist, the quality of reporting and type of population captured varies substantially by data source type and diagnosis method. An intrapartum stillbirth is defined as a fetal death occurring after the onset of labor and prior to delivery. The gold-standard classification of whether a fetus is alive after the onset of labor is the presence of a fetal heartbeat. However, in many contexts, the presence of a fetal heartbeat may not be documented and so diagnosis of the presence of life occurs through other methods postpartum; for example, fetuses who died antepartum can have skin changes consistent with maceration, tissue injury, meconium staining, and edema (\protect\hyperlink{ref-da2016stillbirth}{Da Silva et al. 2016}). However, these types of diagnosis methods are much more prone to misclassification compared to heartbeat diagnoses. In addition, different countries use different definitions of what constitutes a stillbirth, based on gestational age or birthweight thresholds (or a mixture of both). For international comparability, the WHO recommends using the cut-off of 1000g or more at birth (if available), or after 28 completed weeks of gestation (\protect\hyperlink{ref-hug2020neglected}{Hug et al. 2020}). But different countries report stillbirths at different definitions, and definitions may vary subnationally by jurisdiction; for example in the United States, there are eight different definitions by combinations of gestational age and weight (\protect\hyperlink{ref-da2016stillbirth}{Da Silva et al. 2016}). In particular, many countries report stillbirths at 22 weeks gestation. An additional issue stems from differences in data collection systems. While most high-income countries collect information on stillbirths through civil registration and vital statistics (CRVS) systems, many other countries rely on other collection systems that are less reliable and cover only a small portion of the population, and some countries have no information available. As such, a method to estimate the proportion of stillbirths that are intrapartum needs to account for a wide range of data reporting, coverage, and classification issues. However, previous regional estimates of intrapartum stillbirths rely on median proportion-based approaches, making no data adjustments and not accounting for uncertainty in data, estimates, or projections (\protect\hyperlink{ref-hug2020neglected}{Hug et al. 2020}).

To overcome these issues, we formulate a Bayesian hierarchical penalized splines model that takes into consideration different types of data on stillbirth timing from a wide range of reporting and diagnosis systems. The model captures the underlying relationship between changes in the proportion of stillbirths that are intrapartum and the neonatal mortality rate over time, which allows for estimates and projections of the intrapartum proportion to be made even in contexts where available data are limited. Additionally, the model allows for data-driven trends through the use of a penalized splines regression framework. The model is also informed by high-quality auxiliary data to adjust for different gestational age definitions, and allows for varying amounts of uncertainty around data from different sources. We also propose a post-estimation weighting scheme to account for varying levels of coverage in the observed data.

The remainder of the paper is structured as follows. Firstly, we describe data sources and availability, before introducing the modeling framework and presenting some results and validation. We conclude with a discussion of possible extensions to the model.

\hypertarget{data}{%
\section{Data}\label{data}}

Data on the number of stillbirths by timing (intrapartum versus antepartum) can come from a number of different sources. Data from civil registration and vital statistics (CRVS) systems generally has national coverage and are assumed to be relatively high quality compared to other sources. Data on stillbirth timing also come from health management information systems (HMIS) (such as DHIS2), which are commonly used in low- and middle-income countries to record information from a number of health facilities (\protect\hyperlink{ref-moxon2015count}{Moxon et al. 2015}). Subnational data are sourced from either health facility data or population-based studies, most notably through the Global Network Study, which collects data on stillbirths and neonatal deaths in multiple communities across several different countries (\protect\hyperlink{ref-froen2016stillbirths}{Frøen et al. 2016}). Data were extracted and collated through a number of channels, including web-based searches of national statistics' offices, UNICEF country consultations, and literature searches.

Overall, a least one observation of the number of stillbirths that occurred intrapartum and antepartum was available for 92 countries across the period 2000--2020. Table \ref{tab:data-availability} shows the breakdown of data availability by Sustainable Development Goal (SDG) region, suggesting that the most observations are available in the high-income country group.

\begin{table}[H]

\caption{\label{tab:data-availability}Data availability by SDG region}
\centering
\begin{tabular}[t]{lrrr}
\toprule
SDG region & Observations & Countries & Country-years\\
\midrule
Central and Southern Asia & 163 & 7 & 65\\
Eastern and South-Eastern Asia & 57 & 8 & 53\\
Latin America and the Caribbean & 272 & 13 & 171\\
North America, Europe, Australia and New Zealand & 460 & 30 & 368\\
Northern Africa and Western Asia & 43 & 8 & 42\\
Oceania (exc. Australia and New Zealand) & 1 & 1 & 1\\
Sub-Saharan Africa & 280 & 25 & 158\\
\bottomrule
\end{tabular}
\end{table}

Looking at the proportion of observations by data collection system (Table \ref{tab:prop-source}) shows marked differences across the SDG regions. In particular, while 98\% of data in North America, Europe, Australia and New Zealand come from CRVS systems, less than 10\% of the data from Central and Southern Asia do, and almost no data in Sub-Saharan Africa are from CRVS. The majority of data in Central and Southern Asia are from subnational population-based studies. Northern Africa and Western Asia have notable shares from both CRVS and HMIS systems, while data from Sub-Saharan Africa are most from subnational population-based studies or HMIS.

\begin{table}[H]

\caption{\label{tab:prop-source}Data source by SDG region}
\centering
\begin{tabular}[t]{>{\raggedright\arraybackslash}p{6cm}rrrr}
\toprule
SDG region & CRVS & Health facility & Subnat pop-based & HMIS\\
\midrule
Central and Southern Asia & 0.067 & 0.110 & 0.810 & 0.012\\
Eastern and South-Eastern Asia & 0.684 & 0.281 & 0.035 & 0.000\\
Latin America and the Caribbean & 0.794 & 0.044 & 0.162 & 0.000\\
North America, Europe, Australia and New Zealand & 0.980 & 0.015 & 0.004 & 0.000\\
Northern Africa and Western Asia & 0.465 & 0.047 & 0.093 & 0.395\\
Oceania (exc. Australia and New Zealand) & 0.000 & 1.000 & 0.000 & 0.000\\
Sub-Saharan Africa & 0.007 & 0.139 & 0.375 & 0.479\\
\bottomrule
\end{tabular}
\end{table}

Table \ref{tab:prop-def} shows the proportion of observations by SDG based on gestational age or birthweight definitions. The early stillbirth definition refers to either stillbirth at 22 weeks or later, and/or a birthweight of 500g or above. The late stillbirth definition refers to either stillbirth at 28 weeks or later, and/or a birthweight of 1000g or above. For the purposes of our estimation efforts, we are interested in estimating stillbirth timing for late stillbirths, and so definitional differences need to be adjusted for in the modeling framework. As shown in Table \ref{tab:prop-def}, observations based on the early definition make up at least 17\% of all observations across all regions. In addition, observations where it is unclear what definition to use, or where all stillbirths are included, make up a non-negligible proportion in Latin America and the Caribbean, and Sub-Saharan Africa.

\begin{table}[H]

\caption{\label{tab:prop-def}Stillbirth definition by SDG region}
\centering
\begin{tabular}[t]{lrrr}
\toprule
SDG region & early & late & not defined or all\\
\midrule
Central and Southern Asia & 0.429 & 0.534 & 0.037\\
Eastern and South-Eastern Asia & 0.228 & 0.772 & 0.000\\
Latin America and the Caribbean & 0.176 & 0.570 & 0.254\\
North America, Europe, Australia and New Zealand & 0.465 & 0.526 & 0.009\\
Northern Africa and Western Asia & 0.628 & 0.326 & 0.047\\
Oceania (exc. Australia and New Zealand) & 1.000 & 0.000 & 0.000\\
Sub-Saharan Africa & 0.254 & 0.621 & 0.125\\
\bottomrule
\end{tabular}
\end{table}

\hypertarget{additional-data-sources}{%
\subsection{Additional data sources}\label{additional-data-sources}}

In addition to data on the number of intrapartum and antepartum stillbirths, we use several other data sources and estimates to help inform the model and calculate the eventual proportion of stillbirths that are intrapartum (IPSB) at various regional and global aggregations.

Firstly, we use estimates of the neonatal mortality rate (NMR), which is the number of deaths in the neonatal period (first 28 days of life) per 1000 live births, for every country and year in the estimation period of interest. NMR estimates are produced by the UN Interagency Group on Mortality Estimation (UN IGME) as part of annual SDG reporting efforts. Details of the estimates and estimation can be found in UN IGME (\protect\hyperlink{ref-unigme}{2021}). In particular, we use the full posterior samples of NMR for each country-year in order to reflect and propagate the uncertainty in the estimation process.

We also draw upon estimates of the total stillbirth rate (SBR), which is defined as the number of babies born with no sign of life at 28 weeks or more of gestation, per 1,000 total births. These estimates are used to quantify the coverage of data sources within a country, and to weight country-level estimates of IPSB to get regional estimates of IPSB. Similarly to the NMR, we use UN IGME estimates (\protect\hyperlink{ref-hug2020neglected}{Hug et al. 2020}) and use the full posterior samples to better propagate the uncertainty in the SBR estimation process.

As shown in Table \ref{tab:prop-def}, data on stillbirths by timing is available at varying gestational age definitions. In order to inform gestational age adjustments in the estimation process, we use data from Euro-Peristat (\protect\hyperlink{ref-gissler2010perinatal}{Gissler et al. 2010}). The data provided contain intrapartum and antepartum stillbirth counts in 2015 for 17 high-income European countries corresponding to both types of gestational age definitions. For confidentiality reasons, country names in these data are not given. Additionally, to help inform the gestational age adjustment for lower income countries, we draw on data from the Global Health Network (\protect\hyperlink{ref-froen2016stillbirths}{Frøen et al. 2016}), which gives information on stillbirths by timing for eight low- and middle-income countries\footnote{Argentina, Democratic Republic of Congo, Zambia, Guatemala, Bangladesh, India, Pakistan, and Kenya.} at both early and late gestational age definitions.

\hypertarget{model-framework}{%
\section{\texorpdfstring{Model framework \label{section-model}}{Model framework }}\label{model-framework}}

\hypertarget{overview}{%
\subsection{Overview}\label{overview}}

The varying type, quality, and coverage of data on stillbirths by timing that are available across different countries suggests the need for an estimation process that accounts and adjusts for various data characteristics. Firstly, there is a large degree of missingness in many countries across the time period of interest (2000-2021), with many countries only having one observation over the whole period, and more than half of the countries having no observations at all. As such, a hierarchical model, which allows for information exchange across countries within regions may be appropriate for this context. The lack of available data also suggests that one or more covariates may be useful to obtain reasonable trends over time. Secondly, the data that do exist come from a wide range of data collection systems, which are likely to have different levels of coverage of the overall population of interest, and also measurement error in the classification of intrapartum versus antepartum timing. This suggests a need for allowing for different magnitudes of error around observations in a data (or measurement error) model, and also potentially allow for observations to be `down-weighted' if they come from a data system that only covers a small portion of total stillbirths. Finally, data are reported using different definitions (early or late stillbirths). As we expect early stillbirths to have a higher proportion of fetal deaths occurring in the intrapartum period (\protect\hyperlink{ref-getahun2007risk}{Getahun et al. 2007}), the estimation process should account for these definitional issues.

To address these issues, we propose a Bayesian hierarchical penalized splines regression model to estimate the IPSB levels in each country and trends over the period 2000-2021. In this section we describe the model for place-specific IPSB, the weighting process to obtain country-level estimates, and how regional estimates are derived.

\hypertarget{model-for-place-level-ipsb}{%
\subsection{Model for place-level IPSB}\label{model-for-place-level-ipsb}}

We consider data on intrapartum and antepartum stillbirths as available for a specific `place.' In many cases, a `place' refers to data for a whole country (e.g.~data from a nationally-representative CRVS system), but in other contexts, a `place' may refer to a subnational region (such as a state or province) or a single health facility. In our approach, we explicitly model the data at the `place' level, and then reweight the estimates in a second step to obtain country-level estimates.

For data points \(i = 1, \dots, N\), let \(y_i\) and \(z_i\) denote the the number of observed intrapartum and antepartum stillbirths respectively. Then
\begin{equation}
y_i|\phi_i \sim \text{Binomial} (y_i + z_i, \phi_i),
\end{equation}
where \(y_i + z_i\) represents the total number of classified stillbirths\footnote{Note that for some observations, the number of stillbirths of unknown timing was also reported. For the purposes of estimation, unknown timing stillbirths were excluded.} and the parameter \(\phi_i\) represents the proportion of intrapartum stillbirths. The proportion \(\phi_i\) is modeled
\begin{gather}
\text{logit}(\phi_i) = \mu_i + \varepsilon_i \\
\varepsilon_i \sim \text{Normal}(0,\sigma_{\varepsilon,s[i]}^2)
\end{gather}
where \(\mu_i\) describes the ``true'' inverse-logit transformed proportion for the population, and \(\varepsilon_i\) is an error term. The variance \(\sigma_{\varepsilon, s[i]}^2\) of the error term depends on the type of data system of the observation which are expected to vary in reporting quality. Observations from CRVS systems are expected to be the most reliable, and are assigned \(\sigma_{\varepsilon, s[i]} = 0\). Observations from other types of systems, such as those from subnational health facilities, DHIS/HMIS, or population studies are expected to be less reliable and for these methods the variance terms are estimated from data with half-Normal priors:
\begin{equation}
\sigma_{\varepsilon, s} \;
\begin{cases}
= 0 & \text{if } {s = } \text{ CRVS} \\
\sim \text{Normal}^+(0, 1^2) & \text{if } s = \text{health facility, DHIS/HMIS, population study}
\end{cases}
\end{equation}

The parameter \(\mu_i\) represents the mean for the place-level observation \(i\), and is estimated as
\begin{equation}
\mu_i = \beta_0 + \beta_{r[i]} + \beta_{c[i]} + \beta_{p[i]} + \beta_{\text{NMR}} \log \text{NMR}_{c[i], t[i]} + \eta_{p[i], t[i]} + \gamma_{g[i], m[i]},
\end{equation}
where \(\beta_0\) is a global intercept, \(\beta_{r[i]}\) is a region effect for region \(r\), grouped according to the UN Sustainable Development Goals (SDG) regional groupings, and \(\beta_{c[i]}\) is a country effect for country \(c\). These are modeled as
\begin{gather}
  \beta_0 \sim \text{Normal}(0, 1^2) \\
  \beta_{r} \sim \text{Normal}(0, \sigma_{\beta_r}^2) \\
  \beta_{c} \sim \text{Normal}(0, \sigma_{\beta_c}^2)
\end{gather}
where \(\sigma_{\beta_r}\) and \(\sigma_{\beta_c}\) are variance terms to be estimated.

The parameters \(\beta_{p[i]}\) represent place-level effects which are specific to a population or subnational region \(p\) as specified by descriptions in the data. Sub-populations may vary in size depending on data sources within a country. For instance, one sub-population may be only the births at a specific health facility, while another may be all births at government facilities. We assume
\begin{gather}
  \beta_p \sim \text{Normal} (0, \sigma_{\beta_p}^2),
\end{gather}
where \(\sigma_{\beta_p}\) is a variance term to be estimated. For countries where only one sub-population appears in the data, we set \(\beta_p = 0\) for identifiability with the country effect \(\beta_c\).

\begin{figure}[h!]
\centering
\includegraphics[width = 0.8\textwidth]{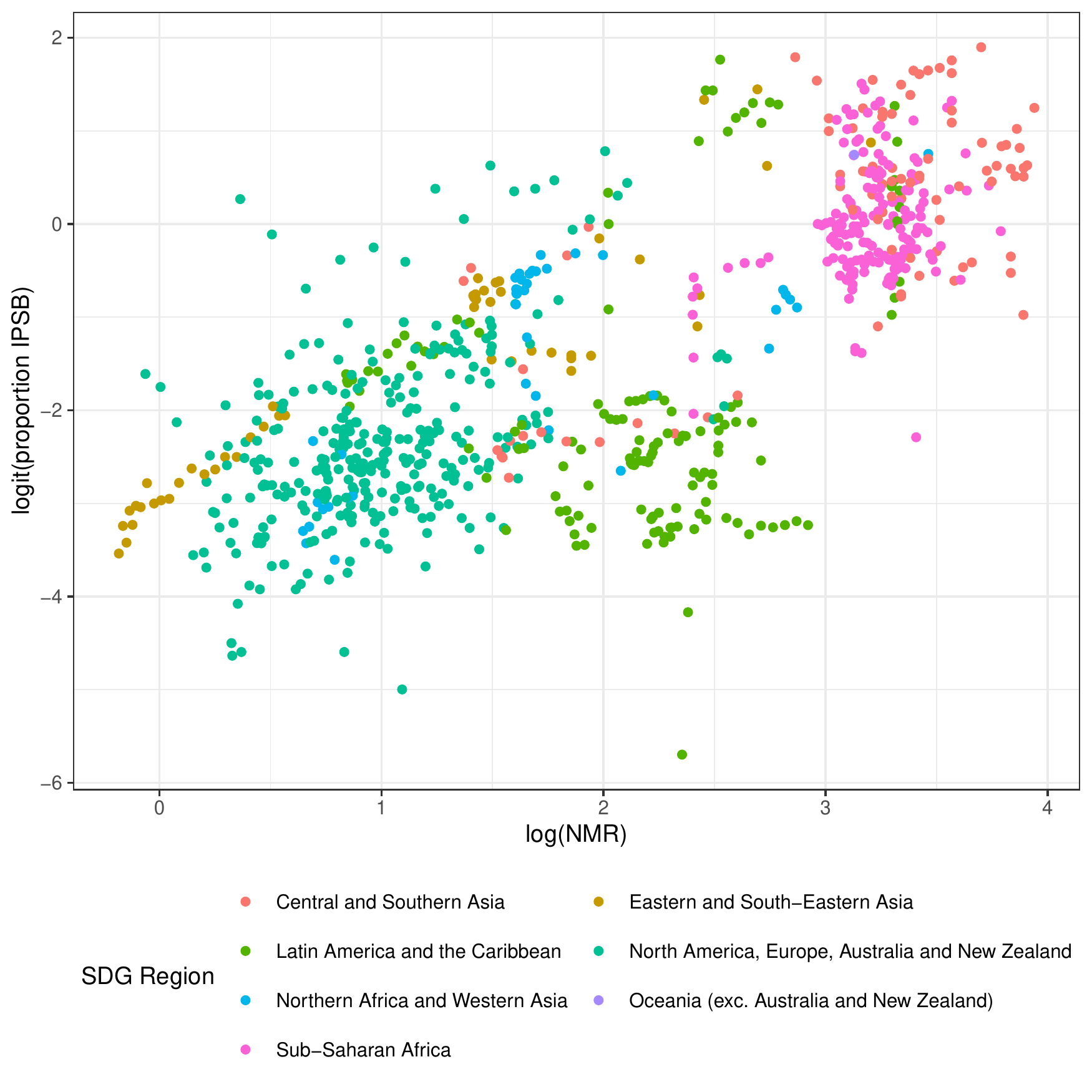}
\caption{Observed proportion of stillbirths that are intrapartum (IPSB) (logit scale) versus log of the neonatal mortality rate. Colors represent different SDG regions.}
\label{fig-nmr}
\end{figure}

We use estimates of the (log) NMR as a covariate in the model. This is motivated by the strong empirical relationship observed between the logit IPSB and log NMR (as shown in Figure \ref{fig-nmr}), and also by the fact that we expect the proportion of stillbirths that are intrapartum to be positively correlated with NMR, as both outcomes are likely to decline in response to improved medical conditions during birth (\protect\hyperlink{ref-joyce2004associations}{Joyce et al. 2004}). The term \(\beta_{\text{NMR}}\) represents the slope of the effect and \(\text{NMR}_{c[i], t[i]}\) refers to the UN IGME point estimate of the national NMR of country \(c\) and time \(t\) of the observation.

To allow for data-driven trends in addition to trends based on changes in country-level NMR, we include a place-time specific component \(\eta_{p, t}\), which is modelled using a first-order penalized splines set up. For each place \(p\), \(\eta_{p, t}\) is defined
\begin{equation}
\eta_{p, t} = \sum_{h=1}^H k_h(t)\alpha_{h, p} 
\end{equation}
where \(k_h(t)\) denotes the \(h^{\text{th}}\) spline function evaluated at time \(t\) and \(\alpha_{h,p}\) denotes its coefficient to be estimated. We use cubic B-splines with knots placed at integer year values. First-order differences in the coefficients, denoted \(\Delta_{h, p}\), are penalized to ensure a level of smoothness in the resulting fit:
\begin{gather}
\Delta_{h, p} = \alpha_{h, p} - \alpha_{h-1,p} \\
\Delta \sim \text{Normal} (0, \sigma_{\Delta}^2)
\end{gather}
where \(\sigma_\Delta\) is an estimated variance term, and the coefficients are constrained to sum to zero
\begin{equation}
\sum_{h=1}^H \alpha_{h,p} = 0.
\end{equation}

Finally, \(\gamma_{g,w}\) is an adjustment for the gestational age definition \(g\) used in the observation (\(g=\) early or \(g=\) late). Separate adjustment factors are estimated for two income groups: \(m =\) high income countries (HIC) and \(m =\) low-middle income countries (LMIC). Since the final desired estimate of the intrapartum stillbirth proportion is with respect to a 28-week gestational age definition of stillbirth, we set \(\gamma_{g=\text{late},m} = 0\) for observations which use a 28-week definition, or comparable ``late'' definition of stillbirth, and estimate adjustment factors for \(\gamma_{g = \text{early}, m}\) to accommodate observations that use a 22-week definition or comparable ``early'' definition.

The adjustment factors are additionally informed by supplementary datasets for which timing-specific stillbirth counts are available under different gestational age definitions. In particular, the high income group is informed by anonymized country-level data from Euro-Peristat. Country names in these data are not available and therefore cannot be joined with the NMR covariate to inform the region or country-level estimates directly. For the low-income group, we incorporate data from the Global Network Maternal Newborn Health Registry, for which timing-specific stillbirth counts according to 22-week and 28-week definitions are typically available. While these latter observations are not anonymized, counts that correspond to the 28 week definition are already used to inform the country intrapartum stillbirth proportions. We therefore model these observations to inform only \(\gamma\) as follows. Let \(\dot{y}_{c,g}\) and \(\dot{z}_{c, g}\) denote respectively intrapartum and antepartum stillbirth counts for some country \(c\) using gestational age definition \(g\). The counts are then modeled
\begin{gather}
  \dot{y}_{c,g}|\rho_{c,g} \sim \text{Binomial} (\dot{y}_{c,g} + \dot{z}_{c,g},\, \rho_{c,g}) \\ 
  \text{logit}\rho_{c,g} = \nu_{c,g} + \gamma_{g, m[c]} \\
  \nu_{c,g} \sim \text{Normal}(0, 10^2)
\end{gather}
where \(\nu_{c,g}\) is given a vague prior and represents the mean under the late gestational age definition. The difference between the proportions in the early and late definitions is therefore captured by the adjustment factor \(\gamma_{\text{early}, m[c]}\), where \(m[c]\) denotes the income group of country \(c\).

For high-income countries \(m =\) HIC, we assign \(\gamma_{g=\text{early}, m =\text{HIC}}\) the prior
\begin{equation}
\gamma_{g=\text{early}, m =\text{HIC}} \sim \text{Normal} (0, 1^2).
\end{equation}
We center the prior adjustment factor for the low-middle income group on the estimate for the high income group,
\begin{equation}
\gamma_{g=\text{early}, m=\text{LMIC}} \sim \text{Normal} (\gamma_{g=\text{early}, m =\text{HIC}} , \sigma_\gamma^2)
\end{equation}

All standard deviation terms to be estimated, \(\sigma_{\beta_r}\), \(\sigma_{\beta_c}\), \(\sigma_{\beta_p}\), \(\sigma_\Delta\), and \(\sigma_{\gamma}\), are assigned a half-Normal(0, 1) prior.

\hypertarget{weighting-adjustment-for-country-level-ipsb}{%
\subsection{Weighting adjustment for country-level IPSB}\label{weighting-adjustment-for-country-level-ipsb}}

\hypertarget{overview-1}{%
\subsubsection{Overview}\label{overview-1}}

Since observations may only pertain to a specific context or geography (e.g.~certain health facilities or subnational region), we do not assume that patterns in the observations generalize to parts of the full national population. We therefore apply a post-estimation weighting step to construct a national estimate as a weighted average of place-level estimates.

Ideally, data pertaining to some place \(p\) would cover the entire country, and capture timing information about all known stillbirths. In this case, the country-level estimate of the IPSB, \({\hat{\phi}}_{c,t}\), would just equal the place-level estimate \({\hat{\phi}}_{p,t}\). In the case where the the entire population of stillbirths is instead reported across multiple places, we would like to know the true proportion \(w_p\) of the country's stillbirths belonging to each place to use as weights. In such a scenario a reasonable estimate of the national rate \(\phi_{c,t}\) would simply be those weights applied to the place-level rates,
\begin{equation}
\hat{\phi}_{c,t} = \sum_{p:c(p) = c} w_p \hat{\phi}_{p,t}.
\end{equation}
In practice however, the observed sub-populations are not in general exhaustive (i.e.~\(\sum w_p < 1\)), meaning that there are stillbirths that are not captured in any of the data sources for the country. To account for this remainder, we add an ``unobserved'' component so that the final estimate is
\begin{equation}
\hat{\phi}_{c,t} = \sum_{p:c[p] = c} w_p \hat{\phi}^\text{obs}_{p,t} + \left(1 - \sum_{p:c[p] =c} w_p\right) \hat{\phi}_{c, t}^\text{unobs}.
\end{equation}
Here, \(\hat{\phi}^\text{obs}_{p,t}\) is the estimate specific to place \(p\), that is, using the estimate of its intercept and time trend. To construct \(\hat{\phi}_{c,t}^\text{unobs}\) we assume a generic place in country \(c\) with unknown intercept and time trend. This assumes that the unobserved stillbirths are centered at the prediction based on its SDG region, NMR level, and estimated country intercept, with additional uncertainty based on the estimated between-sub-population variation and temporal variation. Details on these components are given in the following sections.

\hypertarget{construction-of-weights}{%
\subsubsection{Construction of weights}\label{construction-of-weights}}

In practice, the true weights \(w_p\) are unknown. We instead construct an estimate \(\hat{w}_p\) as the ratio of the number of observed classified stillbirths in place \(p\) to the number of total stillbirths expected nationally, based on UN IGME estimates of overall stillbirths. Under this setup, the estimates for countries at the extremes of data quality are similar to those which would arise from a more conventional hierarchical model where data are modeled at the country level. For instance, if a country reports a single data source with full coverage (e.g.~high quality CRVS), then there is only one place \(p\) (which accounts for all stillbirths in the country), so \(w_p=1\), and the national estimate \(\hat{\phi}_{c,t}\) is simply the estimate \(\hat{\phi}_{p,t}\) informed by that data source. On the other hand, if a country has no data, then the entire estimate consists of \(\hat{\phi}_{c,t}^\text{unobs}\), and is informed entirely by the region and NMR.

To construct the weights for the observed place components, first let \(s_i = y_i + z_i\) denote the sum of observed stillbirths classified as intrapartum or antepartum in observation \(n\). Let \(\tilde{S}_i = \tilde{S}_{c[i], t[i]}\) denote the estimate of total stillbirths from the UN IGME total stillbirth rate model in the country \(c\) and year \(t\) corresponding to the observation. To reflect uncertainty in the number of stillbirths, we directly use posterior samples of \(\tilde{S}_i\) when computing our own posterior samples.

For a place \(p\), we construct its weight \(\hat{w}_p\) as ratio of the sum of observed classified stillbirths in that sub-population, \(\sum_{i:p[i] = p} s_i\), to the sum of estimated stillbirths nationally in the country-years of those observations \(\sum_{i:p[i] = p} \tilde{S}_i\):
\begin{equation}
\hat{w}_p = \frac{\sum_{i:p[i]=p} s_i }{ \sum_{i:p[i] = p} \tilde{S}_i}.
\end{equation}
Terms in the denominator are scaled proportionally for partial year observations. For example, if an data point refers only to one third of the year 2016, then we only add \(S_{c[i], t[i] = 2016}/3\) to the denominator for this observation.

For a country \(c\), we sum the weights of its observed sub-populations to give the weight given to the observed population \(\hat{w}_c = \sum_{p:c[p] = c} \hat{w}_p\). If this quantity exceeds 1, we downscale the weights to add to 1. The weight assigned to the unobserved portion is the remainder \(1-\hat{w}_c\).

The final estimate of the intrapartum proportion \(\hat{\phi}_{c,t}\) for a country is a weighting of observed sub-population components \(\hat{\phi}_{p, t}^\text{obs}\) and an unobserved component \(\hat{\phi}_{c,t}^\text{unobs}\):
\begin{equation}
  \hat{\phi}_{c, t} = \left(\sum_{p:c[p] = c} \hat{w}_p \hat{\phi}_{p,t}^\text{obs} \right) + (1 - \hat{w}_c) \hat{\phi}_{c,t}^\text{unobs}.
\end{equation}

\hypertarget{estimate-for-observed-component}{%
\subsubsection{Estimate for observed component}\label{estimate-for-observed-component}}

The sub-population weights \(w_p\) are applied to sub-population estimates, which are calculated from the estimated parameters,
\begin{gather}
\hat{\mu}_{p,t}^\text{obs} = \hat{\beta}_0 + \hat{\beta}_{r[c[p]])} + \hat{\beta}_{c[p]} +  \hat{\beta}_{p} + \hat{\beta}_{\text{NMR}} \log \tilde{\text{NMR}}_{c[p], t} + \hat{\eta}_{p, t} \\
  \hat{\phi}_{p, t}^\text{obs} = \text{logit}^{-1} (\hat{\mu}_{p,t}^\text{obs}).
\end{gather}

We directly use posterior draws of the country-year neonatal mortality estimates from UN IGME, denoted \(\tilde{\text{NMR}_{c,t}}\), to incorporate appropriate uncertainty about the covariate.

\hypertarget{estimate-for-unobserved-component}{%
\subsubsection{Estimate for unobserved component}\label{estimate-for-unobserved-component}}

The ``unobserved'' component for a country is centered at the estimate given its region and country intercepts and NMR level:
\begin{gather}
  \hat{\mu}_{c,t}^\text{unobs}  = \hat{\beta}_0 +  \hat{\beta}_{r[c]} + \hat{\beta}_c + \tilde{\beta}_{p_c}  + \hat{\beta}_{\text{NMR}} \log \tilde{\text{NMR}}_{c, t} + \tilde{\eta}_{c,t} \\
  \hat{\phi}_{c, t}^\text{unobs} = \text{logit}^{-1} (\hat{\mu}_{c,t}^\text{unobs}) 
\end{gather}
where \(\hat{\beta}_0\), \(\hat{\beta}_{r[c]}\), and \(\hat{\beta}_{\text{NMR}}\) are the usual parameters estimated from data, whereas \(\tilde{\beta}_{p_c}\) and \(\tilde{\eta}_{c,t}\) are new realizations of the sub-population effect and time trend to reflect appropriate uncertainty about the unobserved population.

We assume that the unobserved component consists of a single sub-population. For the place effect \(\beta_{p}\), we use a new realization denoted \(\tilde{\beta}_{p_c}\),
\begin{equation}
\tilde{\beta}_c \overset{RNG}{\sim} \text{Normal} (0, \hat{\sigma}_{\beta_p}^2).
\end{equation}
The new realization of the time trend, denoted \(\tilde{\eta}_{c,t}\), is generated according to the distribution
\begin{gather}
\tilde\eta_{p, t} = \sum_{h=1}^H k_h(t)\tilde{\alpha}_{h, c}  \\
\tilde\Delta_{h, c} = \tilde\alpha_{h, c} - \tilde\alpha_{h-1, c} \\
\tilde{\Delta}_c \overset{RNG}{\sim} \text{Normal} (0, \hat{\sigma}_{\Delta}^2).
\end{gather}

\hypertarget{obtaining-regional-level-estimates}{%
\subsection{Obtaining regional-level estimates}\label{obtaining-regional-level-estimates}}

The intrapartum stillbirth proportion for a region is obtained by multiplying the country-level intrapartum stillbirth proportions by total stillbirth counts, then aggregating at the region level and recalculating the proportion. In particular, if \(\hat{\phi}_{c,t}\) denotes the estimate for country \(c\) at time \(t\), and \(\tilde{S}_{c,t}\) denotes the corresponding total stillbirth count, then \(\hat{\phi}_{c,t} \tilde{S}_{c,t}\) is our estimate of the number of intrapartum stillbirths. We obtain the intrapartum stillbirth proportion for a region \(r\) by calculating
\begin{equation}
\hat\phi_{r,t} = \frac{\sum_{c:r[c] = r} \hat{\phi}_{c, t} \tilde{S}_{c,t}}{\sum_{c:r[c]=r} \tilde{S}_{c,t}},
\end{equation}
where the numerator represents the sum of intrapartum stillbirths in countries in region \(r\), and the denominator represents all stillbirths in countries in region \(r\). Here we again directly use posterior samples of the total stillbirth counts \(\tilde{S}_{c,t}\) when computing our own posterior samples.

\hypertarget{computation}{%
\subsection{Computation}\label{computation}}

We obtain samples of posterior distributions of parameters using Hamiltonian Monte Carlo (HMC), a type of Markov Chain Monte Carlo sampling, implemented in Stan using the \texttt{cmdstanr} interface (\protect\hyperlink{ref-cmdstanr2022}{Gabry \& Češnovar 2022}). For each of 4 HMC chains, we perform 1000 warmup iterations before taking the next 1000 post-warmup iterations as our posterior samples. We monitor for convergence using standard diagnostic measures. R-hat (\(\hat{R}\)) values for all parameters are under 1.02 (\protect\hyperlink{ref-stan2019stan}{Stan Development Team 2019}), and visual inspection of traceplots do not indicate sampling pathologies.

\hypertarget{results}{%
\section{Results}\label{results}}

In this section we illustrate regional estimates of the proportion of stillbirths that are intrapartum (IPSB) over the period of interest. Additionally, we demonstrate the impact of several key components of the modeling process on country-level estimates of IPSB.

\begin{figure}[h!]
\centering
\includegraphics[width = 0.8\textwidth]{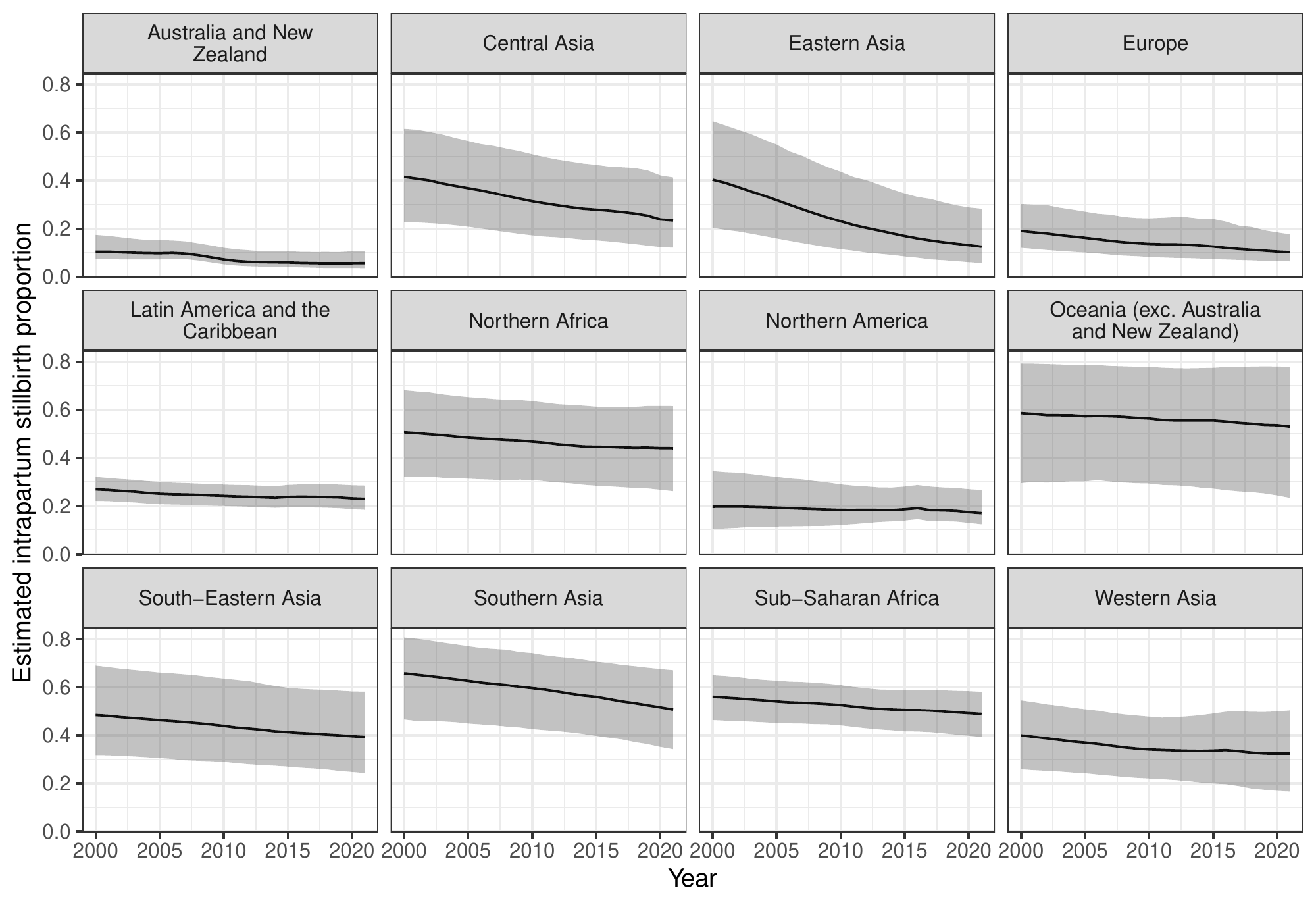}
\caption{Regional estimates of the proportion of stillbirths that are intrapartum.}
\label{fig-sdg}
\end{figure}

Figure \ref{fig-sdg} shows regional estimates of IPSB, with the shaded areas representing 90\% credible intervals. There is substantial regional variation in the level and trends of IPSB. Australia and New Zealand, and Europe, have low and declining IPSB of around 10\%, while regions such as Northern Africa and Sub-Saharan Africa have IPSBs of 40-50\% and show little evidence of decline at the regional level over the time period. Eastern Asia shows evidence of the most rapid decline since 2000, falling from around 40\% to 12\% over the 20-year period. The largest uncertainty in estimates is in the Oceania region, reflecting the lack of available data in this region.

Figure \ref{fig-sdg} shows the impact of the inclusion of the gestational age adjustment in the model. Country `A' shown only has data available on the timing of stillbirths based on the early (22 weeks gestation) period. Estimates without the gestational age adjustment (shown in blue on the right hand side of the graph) tend to follow the data closely. With the addition of the adjustment in the model, the final estimates (shown in red on the left hand side) are shifted downwards, to reflect the fact that stillbirths following the late (28 weeks) definition are likely to have a lower IPSB. Note that the the posterior median estimate for the gestational age adjustment \(\gamma_{g=\text{early},m}\) in high-income countries is 0.31 (90\% CI: {[}0.29, 0.33{]}).

\begin{figure}[h!]
\centering
\includegraphics[width = 0.8\textwidth]{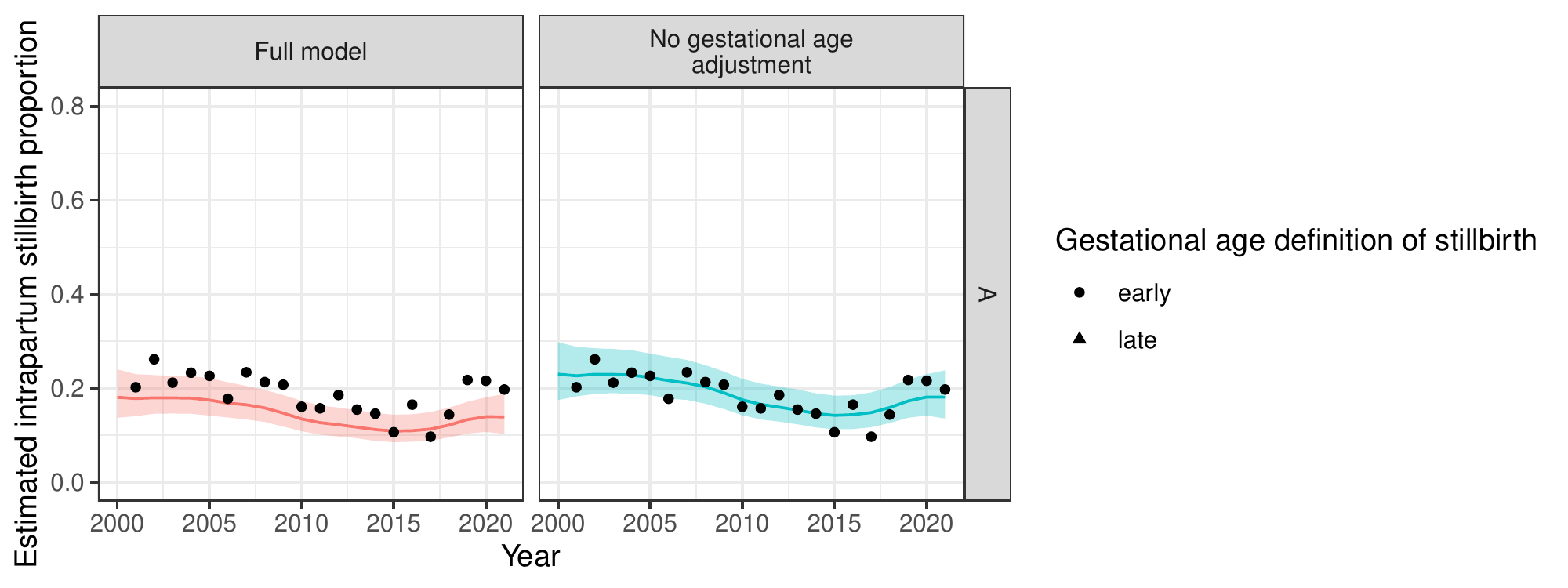}
\caption{Effect of gestational age adjustment on final estimates for country `A'.}
\label{fig-gest}
\end{figure}

\begin{figure}[h!]
\centering
\includegraphics[width = 0.8\textwidth]{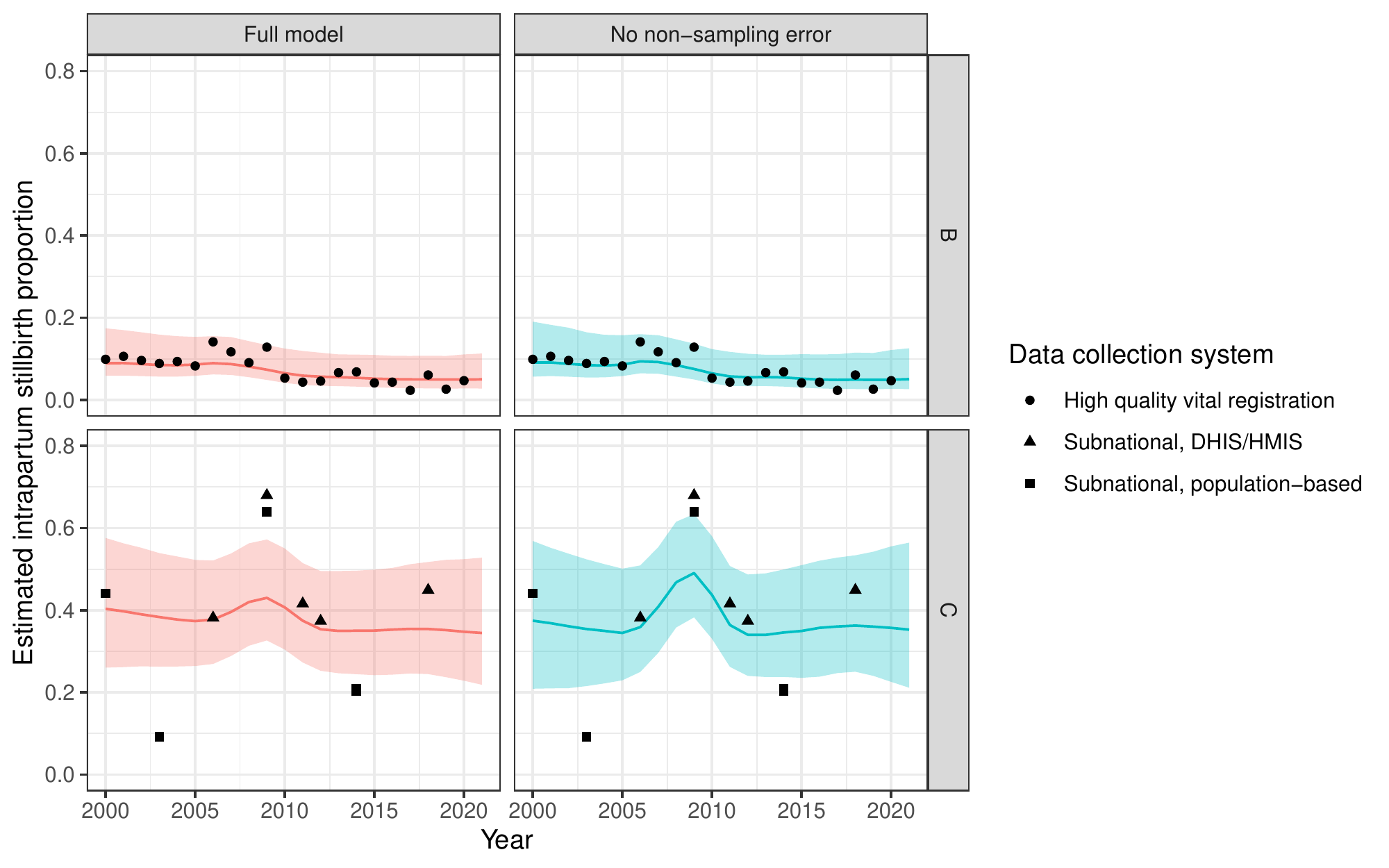}
\caption{Effect of allowing for non-sampling error to vary by source type on final estimates for country `B' (which has vital registration data) and country `C' (which has a mix of subnational data available). Points that report on the same place or sub-population are connected by lines.}
\label{fig-nse}
\end{figure}

In Figure \ref{fig-nse}, we illustrate the effect on estimates of allowing non-sampling error to vary by data source type for two countries with different types of data available. In the figure, estimates shown in blue in the right-hand column are based on a model with no non-sampling error estimated, while estimates shown in red in the left-hand side are the final estimates from the full model. For Country `B,' which has data from a national high-quality vital registration system, there is no effect of non-sampling error on the estimates, as this type of error is assumed to be zero for such data sources. In contrast, Country `C' has data from several subnational sources, both from HMIS and population-based studies. As these subnational sources have additional error that is estimated to be non-zero, the resulting estimates over time are smoother, and in particular are less influenced by the particularly high observations around 2009.

\begin{figure}[h!]
\centering
\includegraphics[width = 0.8\textwidth]{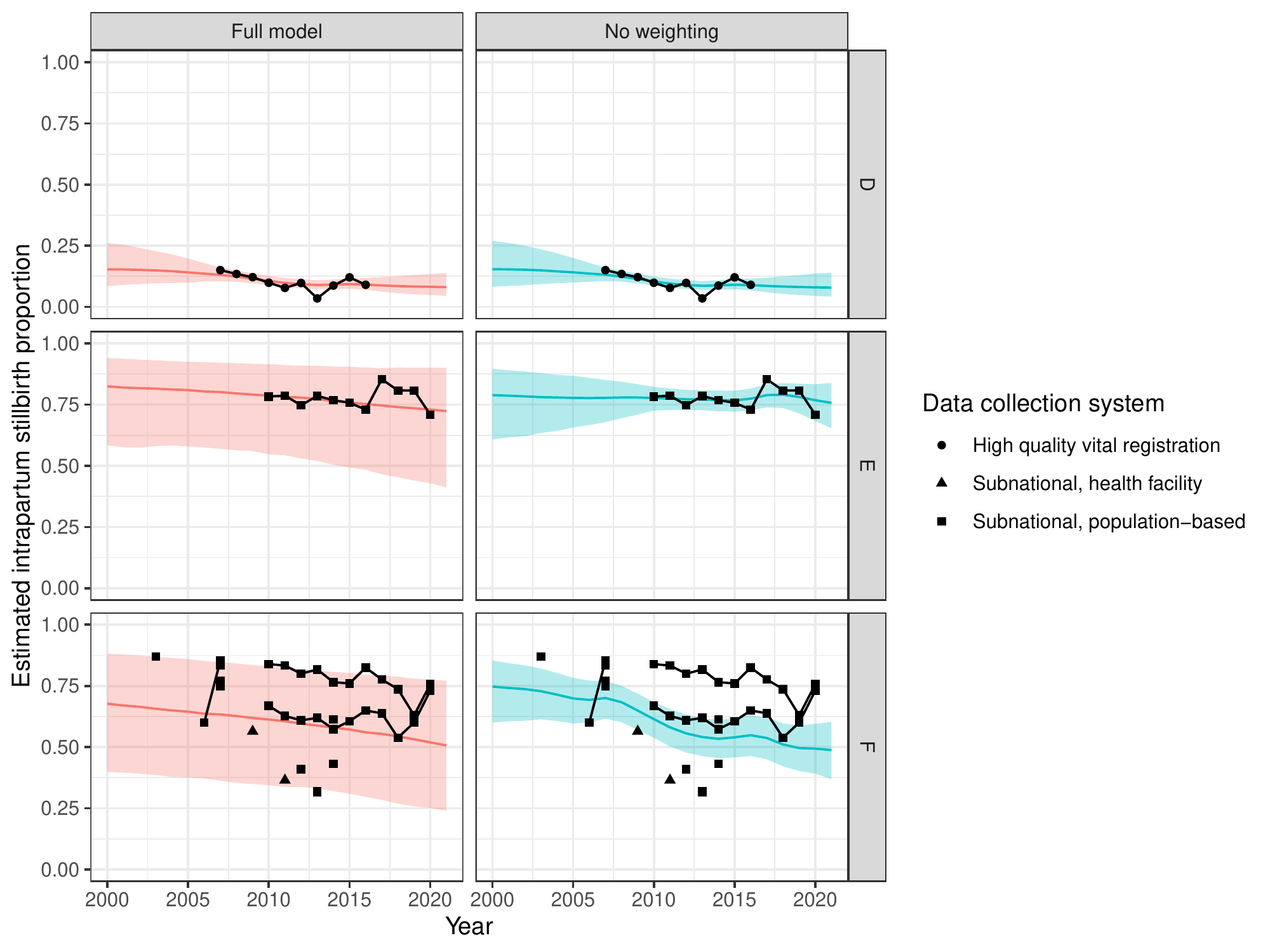}
\caption{Effect of weighting scheme on final estimates, for countries with a range of different data availability situations.}
\label{fig-weighting}
\end{figure}

Finally we illustrate the impact of the post-estimation weighting scheme on country-level estimates for three countries with different data availability situations (Figure \ref{fig-weighting}). In the ``No weighting'' setup on the right, we use a more conventional hierarchical setup in which the \(\beta_p\) terms are still estimated when there are multiple data sources, but are not included when presenting the national estimates. Instead of place-level time trends, a single national time trend is shared across sub-populations.

Country `D' has data that come from a CRVS system and capture all stillbirths in the country. As such, the estimated weights for that data series are approximately 1 and the post-estimation weighting scheme has no effect on the final estimates. In contrast Countries `E' and `F' have one or more subnational data sources, which capture well below the estimated total number of stillbirths. The final estimates that include the re-weighting step thus indicate a higher level of uncertainty around the IPSB levels.

\hypertarget{validation}{%
\subsection{Validation}\label{validation}}

We evaluated the model using two out-of-sample prediction exercises. In the first exercise we fit the model using observations before 2017 as the training set (which represent 78\% of all observations), leaving out observations from 2017 and later to use as a test set. In the second exercise we perform a 10-fold cross-validation. Results from the validation exercises are given below. We calculate mean absolute error as the average of the absolute differences between posterior predicted medians and observed IPSB proportions in held-out data points, and prediction interval coverage as the proportion of observed IPSB proportions which fall within their respective posterior prediction interval.

Table \ref{tab:val-recent20} shows validation metrics from the first prediction exercise, and Table \ref{tab:val-10foldcv} shows validation metrics from 10-fold cross validation. Mean absolute error in both validation exercises is under 5\%, and overall prediction interval coverage in both cases is close to, but slightly lower than nominal, which suggests a slight underestimation of the variability in the data.

There are some apparent regional differences in the performance of the model. In the Oceania region, there is only one data point in 2018 which also represents the only available observation in that country, resulting in high error in prediction in this region. In the Northern Africa and Western Asia region, prediction coverage is low, possibly due to relatively high between- and within-country variability. Particularly poor performance here in the first validation exercise is partially due to a series of data points for a country in this region which deviate from the NMR trend beyond what is expected from the estimated spline component. Finally, high error in the point estimates for Central and Southern Asia region can be partially explained by a relatively high proportion of low-quality subnational observations which are difficult to predict. The reasonably high prediction interval coverage, however, suggests that variability and uncertainty are well calibrated for observations in this region.

\begin{table}

\caption{\label{tab:val-recent20} Model evaluation metrics using 2000-2016 data as a training set and data from 2017 onward as a test set.}
\centering
\begin{tabular}[t]{lr>{\raggedleft\arraybackslash}p{3cm}}
\toprule
Region & Mean absolute error & 95\% prediction interval coverage\\
\midrule
Global & 0.044 & 0.917\\
Central and Southern Asia & 0.095 & 0.850\\
Eastern and South-Eastern Asia & 0.014 & 1.000\\
Latin America and the Caribbean & 0.028 & 0.960\\
North America, Europe, Australia and New Zealand & 0.028 & 0.922\\
\addlinespace
Northern Africa and Western Asia & 0.049 & 0.615\\
Oceania (exc. Australia and New Zealand) & 0.271 & 1.000\\
Sub-Saharan Africa & 0.052 & 0.979\\
\bottomrule
\end{tabular}
\end{table}

\begin{table}

\caption{\label{tab:val-10foldcv} Model evaluation metrics from 10-fold cross validation.}
\centering
\begin{tabular}[t]{lr>{\raggedleft\arraybackslash}p{3cm}}
\toprule
Region & Mean absolute error & 95\% prediction interval coverage\\
\midrule
Global & 0.041 & 0.925\\
Central and Southern Asia & 0.102 & 0.943\\
Eastern and South-Eastern Asia & 0.039 & 0.980\\
Latin America and the Caribbean & 0.025 & 0.928\\
North America, Europe, Australia and New Zealand & 0.021 & 0.927\\
\addlinespace
Northern Africa and Western Asia & 0.029 & 0.829\\
Oceania (exc. Australia and New Zealand) & 0.271 & 1.000\\
Sub-Saharan Africa & 0.063 & 0.916\\
\bottomrule
\end{tabular}
\end{table}

\newpage

\hypertarget{discussion}{%
\section{Discussion}\label{discussion}}

In this paper we proposed a Bayesian hierarchical penalized splines regression model to estimate the proportion of stillbirths that are intrapartum in all countries worldwide. The model sought to address a multitude of data availability, quality, and coverage issues, which makes obtaining reliable estimates of this indicator particularly challenging in low- and middle-income countries. The proposed model includes a measurement error model to take into account different data collection systems; a gestational age definition adjustment, which is informed by high-quality data across a range of different countries; and hierarchical pooling of information from observations within the same country and region of the world. Trends are informed by both trends in the neonatal mortality rate, and observed data through a penalized splines set up, which allows for temporal patterns to be modeled in data-sparse situations. Validation exercises suggest that the model is reasonably well calibrated.

Results at the regional level suggest substantial variation in the levels and trends in the proportion of stillbirths that are intrapartum globally. While proportions are low and stable in regions such as Australia/New Zealand and Europe (at around 10\%), results suggest that more than half of all stillbirths occur intrapartum in Southern Asia and Sub-Saharan Africa. These high levels coupled with slow declines over time suggest there is still much progress to be made to end preventable stillbirths worldwide.

Our approach treats each series of observations as information about the timing of stillbirths from a particular `place' or sub-population. The coverage of a `place' can range from a whole country, right down to a single health facility in a subnational area. The subsequent country-level estimate of IPSB is then a weighted average of the place-level effects, plus unobserved effects, where the weights are calculated based on the relative share of overall stillbirths observed. In comparison to assuming the every data series is an observation of the IPSB for a whole country, this approach generally leads to larger estimated levels of uncertainty. As the weights are estimated based on the number of observed stillbirths alone, we are allowing larger samples to influence final estimates more heavily than smaller samples. This approach was partly employed to downweigh the influence of non-representative smaller samples, as it is unlikely that `place'-level samples are nationally representative, given that they may only cover government-owned health facilities, or health facilities in a particular city or rural area, for example. In future work, we plan to investigate the construction of these weights in more depth. For instance, in some cases we may have detailed enough information about the source of a subpopulation data series to infer the types of women giving birth that were captured in the data, particularly if the geographic location of the observations is known and can be matched with census or survey data. In these cases, a post-stratification based approach may be possible, to more adequately control for the non-representativeness of different subpopulation samples.

Currently, estimates of stillbirth by timing and the total stillbirth rate are performed separately, with the total stillbirth rate being estimated using a Bayesian sparse regression model with temporal smoothing (\protect\hyperlink{ref-wang2022estimating}{Wang et al. 2022}). As detailed above, we combine our estimates of the intrapartum proportion with posterior samples of total stillbirths to obtain regional estimates and associated uncertainty. Future work could investigate the possibility of estimating both quantities in the same modeling framework. As there is generally more data available for total stillbirth rates, combining the estimation of two approaches may help to inform estimates by timing, and would also help to better incorporate uncertainty from different sources (\protect\hyperlink{ref-schumacher2022flexible}{Schumacher et al. 2022}).

Stillbirths have traditionally been overlooked by demographers and public health professionals, with more emphasis being placed on improving infant and maternal survival outcomes. But as substantial health inequalities persist across regions of the world, there have been calls to monitor stillbirths, working towards the goal of ending all preventable stillbirths. From a population dynamics perspective, there has been increased work in understanding fetal outcomes in conjunction with maternal and infant outcomes, in order to allow for a more complete assessment of mortality conditions and improvement (\protect\hyperlink{ref-hathi2022population}{Hathi 2022}). The estimation work presented here aimed to improve estimates of stillbirths by timing, but by doing so highlighted the vast array of data availability and quality issues in this area. Future resource efforts should not only focus on improving access to essential healthcare during pregnancy and childbirth, but also in helping national stakeholders to improve data systems to collect reliable information in order to help those most in need.

\newpage

\hypertarget{references}{%
\section*{References}\label{references}}
\addcontentsline{toc}{section}{References}

\hypertarget{refs}{}
\begin{CSLReferences}{1}{0}
\leavevmode\vadjust pre{\hypertarget{ref-da2016stillbirth}{}}%
Da Silva FT, Gonik B, McMillan M, Keech C, Dellicour S, Bhange S, Tila M, Harper DM, Woods C, Kawai AT, others (2016) Stillbirth: Case definition and guidelines for data collection, analysis, and presentation of maternal immunization safety data. \emph{Vaccine} \textbf{34}:6057

\leavevmode\vadjust pre{\hypertarget{ref-de2016stillbirths}{}}%
De Bernis L, Kinney MV, Stones W, Hoope-Bender P ten, Vivio D, Leisher SH, Bhutta ZA, Gülmezoglu M, Mathai M, Belizán JM, others (2016) Stillbirths: Ending preventable deaths by 2030. \emph{The lancet} \textbf{387}:703--716

\leavevmode\vadjust pre{\hypertarget{ref-froen2016stillbirths}{}}%
Frøen JF, Friberg IK, Lawn JE, Bhutta ZA, Pattinson RC, Allanson ER, Flenady V, McClure EM, Franco L, Goldenberg RL, others (2016) Stillbirths: Progress and unfinished business. \emph{The Lancet} \textbf{387}:574--586

\leavevmode\vadjust pre{\hypertarget{ref-cmdstanr2022}{}}%
Gabry J, Češnovar R (2022) Cmdstanr: R interface to 'CmdStan'.

\leavevmode\vadjust pre{\hypertarget{ref-getahun2007risk}{}}%
Getahun D, Ananth CV, Kinzler WL (2007) Risk factors for antepartum and intrapartum stillbirth: A population-based study. \emph{American journal of obstetrics and gynecology} \textbf{196}:499--507

\leavevmode\vadjust pre{\hypertarget{ref-gissler2010perinatal}{}}%
Gissler M, Mohangoo AD, Blondel B, Chalmers J, Macfarlane A, Gaizauskiene A, Gatt M, Lack N, Sakkeus L, Zeitlin J (2010) Perinatal health monitoring in europe: Results from the EURO-PERISTAT project. \emph{Informatics for Health and Social Care} \textbf{35}:64--79

\leavevmode\vadjust pre{\hypertarget{ref-hathi2022population}{}}%
Hathi P (2022) Population science implications of the inclusion of stillbirths in demographic estimates of child mortality.

\leavevmode\vadjust pre{\hypertarget{ref-hug2019national}{}}%
Hug L, Alexander M, You D, Alkema L, Child UIG for (2019) National, regional, and global levels and trends in neonatal mortality between 1990 and 2017, with scenario-based projections to 2030: A systematic analysis. \emph{The Lancet Global Health} \textbf{7}:e710--e720

\leavevmode\vadjust pre{\hypertarget{ref-hug2020neglected}{}}%
Hug L, Mishra A, Lee S, You D, Moran A, Strong KL, Cao B (2020) A neglected tragedy the global burden of stillbirths: Report of the UN inter-agency group for child mortality estimation, 2020.

\leavevmode\vadjust pre{\hypertarget{ref-hug2021global}{}}%
Hug L, You D, Blencowe H, Mishra A, Wang Z, Fix MJ, Wakefield J, Moran AC, Gaigbe-Togbe V, Suzuki E, others (2021) Global, regional, and national estimates and trends in stillbirths from 2000 to 2019: A systematic assessment. \emph{The Lancet} \textbf{398}:772--785

\leavevmode\vadjust pre{\hypertarget{ref-joyce2004associations}{}}%
Joyce R, Webb R, Peacock J (2004) Associations between perinatal interventions and hospital stillbirth rates and neonatal mortality. \emph{Archives of Disease in Childhood-Fetal and Neonatal Edition} \textbf{89}:F51--F56

\leavevmode\vadjust pre{\hypertarget{ref-lawn2016stillbirths}{}}%
Lawn JE, Blencowe H, Waiswa P, Amouzou A, Mathers C, Hogan D, Flenady V, Frøen JF, Qureshi ZU, Calderwood C, others (2016) Stillbirths: Rates, risk factors, and acceleration towards 2030. \emph{The Lancet} \textbf{387}:587--603

\leavevmode\vadjust pre{\hypertarget{ref-moxon2015count}{}}%
Moxon SG, Ruysen H, Kerber KJ, Amouzou A, Fournier S, Grove J, Moran AC, Vaz LM, Blencowe H, Conroy N, others (2015) Count every newborn; a measurement improvement roadmap for coverage data. \emph{BMC pregnancy and childbirth} \textbf{15}:1--23

\leavevmode\vadjust pre{\hypertarget{ref-schumacher2022flexible}{}}%
Schumacher AE, McCormick TH, Wakefield J, Chu Y, Perin J, Villavicencio F, Simon N, Liu L (2022) A flexible bayesian framework to estimate age-and cause-specific child mortality over time from sample registration data. \emph{The Annals of Applied Statistics} \textbf{16}:124--143

\leavevmode\vadjust pre{\hypertarget{ref-stan2019stan}{}}%
Stan Development Team (2019) Stan modeling language users guide and reference manual version 2.25.

\leavevmode\vadjust pre{\hypertarget{ref-unigme}{}}%
UN IGME (2021) Levels and trends in child mortality levels and trends in child mortality.

\leavevmode\vadjust pre{\hypertarget{ref-wang2022estimating}{}}%
Wang Z, Fix MJ, Hug L, Mishra A, You D, Blencowe H, Wakefield J, Alkema L (2022) Estimating the stillbirth rate for 195 countries using a bayesian sparse regression model with temporal smoothing. \emph{The Annals of Applied Statistics} \textbf{16}:2101--2121

\leavevmode\vadjust pre{\hypertarget{ref-world2014every}{}}%
World Health Organization (2014) Every newborn: An action plan to end preventable deaths.

\end{CSLReferences}

\end{document}